# A Short Introduction to Metal Forming

*C. Moussa and P. Montmitonnet*
Mines Paris, PSL University, Centre de Mise en Forme des Matériaux (CEMEF), UMR7635 CNRS, 06904 Sophia Antipolis, France

**Abstract**
The present paper is a short introduction to metal forming. The paper highlights the different disciplines that should be considered when analyzing metal forming. Hence, mechanical flow, heat flow, mechanical and thermal contact, and microstructure evolution are presented within the perspective of metal forming. For every aspect, well-established knowledge is presented in addition to scientific open questions on which active research is still done.

## 1  Introduction

The use of metals and alloys started thousands of years ago and marked a milestone in human history. Our ability to shape them allowed us to form tools and to improve productivity. Today, metal forming is still a scientific and technological challenge. For a long time, the main challenge was to shape the alloy without the occurrence of damage or fracture. Today, it is necessary to optimize the induced microstructure to control in-service properties. Additionally, the use of fully or partially recycled alloys, in addition to energy reduction during metal forming, will probably be the main challenge for the next decade.

Metal forming can be done during solidification through molding or through additive manufacturing, through powder sintering, through machining and through shaping in the solid state. The latter, which basically consists in controlled plastic deformation, is the forming approach that this present paper aims to introduce. Generally, a processing route or a transformation sequence is defined with several thermo-mechanical operations that modify the shape and the desired microstructure in terms of phase distribution, crystallographic texture, grain size… The control of processing route is done through the control of several aspects of the process. Some of them can be listed such as the temperature (and its homogeneity), the heating and the cooling rate, the soaking time, the deformation mode (hot or cold), the strain rate, the sequence between deformation and annealing, the total strain per deformation step. In addition, one can list the processes related to the finishing such as thermo-chemical or thermo-mechanical surface treatments, machining and finally assembly. Each one of those has its own scientific and technical challenges. In the aim of reducing costs and production time, it is also necessary to control the roughness of the final piece to avoid any additional finishing steps. Therefore, the materials forming is a high technicity sector, continuously evolving and engaging several scientific disciplines from classical continuum mechanics (specifically plasticity theory) to physical metallurgy, chemistry of solids, tribology, thermal physics, applied mathematics for numerical tools development, and nowadays, data analysis and the use of artificial intelligence.

Several classifications can be made for material forming such as hot vs cold or sheet vs bulk. Thanks to high ductility and low mechanical resistance, hot deformation is generally used for large deformations mainly in the early stages of the processing route. However, it induces poor surface quality (because of massive oxidation) and dimensional tolerance. Therefore, cold deformation is generally used in the last stages of the processing route to improve surface quality and dimensional tolerances.

The present paper is a short introduction to material forming that aims to define the main aspects that should be considered. This choice was made to avoid listing all the industrial processes that are used which would be of less scientific interest. Therefore, the paper focuses on four aspects of material forming: material flow, heat flow, tribology and microstructure as illustrated in Fig. 1a. Additionally, all those aspects interact together leading to a highly coupled problem as illustrated in Fig.1b. In our opinion, controlling those four aspects would allow a successful material forming operation. For a detailed review of different materials forming processes, the reader is invited to read [1] for mechanical aspects, [2] for thermal aspects, [3] for tribological aspects and [4] for the physical metallurgical aspects and the microstructural evolutions that may take place.

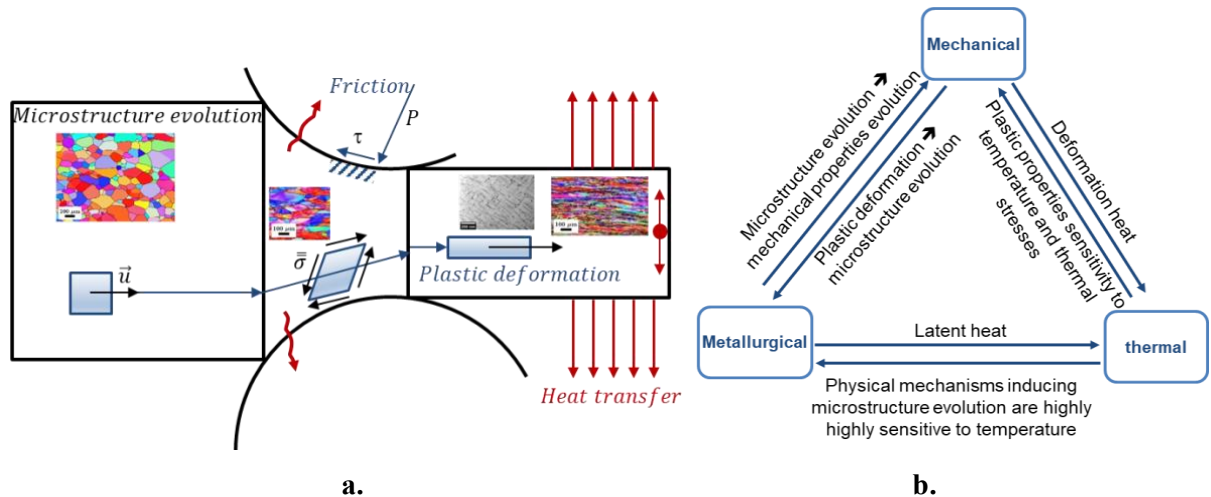

**a.** **b.**
**Fig. 1: a.** Schematic representation of the main aspects involved during material forming: heat flow, material flow, tribology and microstructural evolution through a sheet rolling illustration and **b.** Main multi-physical interactions during plastic deformation (figure adapted from [5]).

## 2 Material flow

### 2.1 Why "flow"?

Confronted to large plastic deformation as imposed in metal forming, it is common practice to refer to "plastic flow", coining an analogy with fluid mechanics. Indeed, it has been noticed early in the development of the plasticity theory that steady state plastic deformation patterns bear some resemblance with fluid streamlines in simple geometries [6].

It is however most uncommon to observe turbulence in solid metal flow, except maybe during explosive welding which often results in a wavy interface – with a shape of waves breaking on a shore. This is when dynamic, acceleration forces overcome the plastic stress itself. Globally, the fluid analogy would therefore be with highly viscous complex fluids, which do not show turbulence. Indeed, FEM modelling of hot metal forming often uses non-linear viscous (viscoplastic) constitutive models [7], very similar to what is used for molten polymer processing.

### 2.2 General features

Another striking characteristic of plasticity, compared with elasticity, is the possibility of strain concentration and localization in shear bands (macroscopic, at the workpiece scale, or microscopic, slip bands within grains). These often result in severe local, heterogeneous plastic heating leading to local microstructural transformations on the one hand, in damage on the other hand, possibly until fracture. But this tendency to localization is exploited in processes like metal cutting or fine blanking where a highly localized deformation results in a "clean" cut surface.

Indeed, in plane strain two-dimensional problems, the equations of plastic deformation result in hyperbolic Partial Differential Equations (PDE), whereas elasticity gives elliptic PDEs. Plasticity is therefore somewhat similar in mathematical structure to shock waves in supersonic aerodynamics. The result is that while *elastic* stress and strain fields extend to infinity, the *plastic* stress and strain state tend to be restricted to a vicinity of the applied loads; beyond this limit, they decay into elastic stress and strain, which extend to infinity.

From this character derives also a powerful semi-analytical solution tool, the Slip Line Field method (SLF), which offers *exact* solutions in simple cases. Although complex to handle, it has allowed a very deep understanding of most forming processes and of the material history therein [8]. This hyperbolic character is lost, strictly speaking, when going to axis-symmetrical or 3D problems, but the learnings from the Plane Strain SLF remain useful for a diversity of problems, from indentation hardness testing to wire drawing or strip rolling processes, orthogonal metal cutting or metal extrusion.

After an elastic loading stage, plasticity begins when a certain combination of the components of the current stress tensor reaches a critical value, the yield stress (by convention, the yield stress in a simple tension test). The mathematical shape of this combination defines the *yield criterion*. The way plastic strain rate (which quantifies the plastic flow) develops then is governed by the *flow rule*. The latter two form the plastic constitutive model. In general, metals harden during strain (at least in cold plastic deformation), this strain hardening model is a third, complementary feature of the plastic behavior.

A general feature of the plasticity of dense metals is its isochoricity, i.e. plasticity develops without volume change; this is more commonly termed "incompressibility". A corollary is that the yield stress is little, if at all, sensitive to hydrostatic pressure, contrary to fluid viscosity e.g. This is quite different from e.g. soil mechanics or powder metal forming (compaction + sintering, Hot Isostatic Pressing (HIP) etc.), where plastic flow results in volume mass variation: densification under hydrostatic stress, de-densification under shear e.g. Incompressibility comes from the physical origin of the plastic deformation of dense metals, i.e. dislocations glide on their slip systems: this microscopic process is sensitive to the resolved shear stress on the slip planes, independent from hydrostatic pressure. Of course, this might not be absolutely true at very high pressures (some GPa), but 99% of dense metal forming modelling (rolling, forging, drawing, cutting etc.) is righteously performed with incompressible plastic constitutive models.

Other important features include:

- The yield stresses in tension and compression are generally equal; this is not systematically true but remains a common and useful approximation;

- the yield surface, i.e. the geometrical representation of the yield criterion function in the 6-dimensional space of the independent stress components (which can be reduced to 3 dimensions by considering principal stresses), must be convex to ensure positive energy dissipation;

- for dense metals, the flow rule states that the strain rate tensor is born by the normal to the yield surface in the same space. This can be proved considering again dislocation glide mechanisms. This implies that the strain rate tensor and the stress *deviator* are proportional.

Considering all this, one may derive in a fairly simple way the most common plastic constitutive models [7]. Smooth criteria (everywhere differentiable yield surfaces) include von Mises' yield criterion and associated flow rule for isotropic plasticity (i.e. the yield stress e.g. in tension is the same in all material directions). In principal stress state, the yield surface is an infinite cylinder of circular cross section and axis necessarily in the (1,1,1) direction due to incompressibility; in principal *deviatoric* stress space, it becomes a sphere. It can be generalized into the Hill's criterion for anisotropic plasticity, a very common situation in reality; taking into account the difference of yield stress in tension between e.g. the rolling direction (RD) and the transverse direction (TD), the section of the cylinder becomes an ellipse, the shape of the yield surface in deviatoric space is an ellipsoid. Hill's criterion is often a rather crude approximation which misses important features of plastic anisotropy. For this reason, a lot of more complex models have been developed [9] and this remains a very active research field. Finally, the non-smooth Tresca's yield criterion is represented by a yield surface in the form of an infinite cylinder with a hexagonal cross section along axis (1,1,1). The corners of the hexagon form vertices which favor plastic localization, but such singularities are difficult to handle. Indeed, the normal to the yield criterion, which bears the strain rate tensor direction, becomes undefined at vertices, which requires a specific treatment when the stress state approaches these corners. This is why this criterion is not very popular for numerical simulation.

## 2.3 Solution methods

The SLF method has already been cited for simple geometries in 2D Plane Strain. Owing to its complexity, it is now practically abandoned, except existing, exact solutions used as references. Bound methods (velocity field-based Upper Bound, stress field-based Lower Bound) have been very useful for practical applications [10] but are now practically forgotten.

The Slab Method for 2D Plane Strain strip rolling models is based on stress tensor approximations and results in a 1D problem (stress depends on the RD coordinate only) so that a single Ordinary Differential Equation has to be solved. It is still used extensively in cold strip rolling, where the approximations made are reasonable; it is the standard in the industrial context, workshop as well as R&D [11]. It can be generalized to take into account the stress variations along TD also, which are quite important for certain practical problems (roll deformation, profile and flatness defects…). In this case, the Finite Difference Method can be viewed as more general and more practical [12].

But considering all processes, where Plane Strain cannot be invoked, the only really universal method today is the Finite Element Method (FEM). It has been developed since the 1970s, in 2D first [13] then in 3D [14]-[16] as the computer powers made it possible. As large distortions are involved, automatic remeshing was a key step in their industrial development [17]. Today, one may say that all processes can be handled with the FEM, including solidification-based Additive Manufacturing [18].

Automatic process optimization is now well established [19]. It consists in embedding FEM computation in a cost-function minimization loop, gradient-based or, more commonly now, Genetic Algorithm or Evolutionary Method-based.

## 2.4 Constitutive model

A major stumbling stone is the precision of constitutive equations. Anisotropy e.g. remains a real difficulty, in particular for sheet metal forming where in-plane anisotropy (difference between RD and TD) as well as normal anisotropy (preferential thinning vs preferential in-plane elongation) are crucial in processes where, generally, thickness is not imposed by the tooling. Local thinning, necking (thickness strain localization) and eventually fracture are therefore controlled by fine features of the constitutive model (Lankford coefficient [20]). These are not so easy to measure under conditions representative of the process (high triaxiality) while being crucial for key industrial problems. The prediction of residual stress (springback, delayed fracture…) is also very sensitive to such details of the plastic (and elastic) behavior. As stated before, efforts are going on to develop and validate more general and precise plastic anisotropy models.

In hot forming, a major difficulty lies in the microstructural transformations taking place during deformation (recovery = dislocation annihilation, dynamic recrystallization = nucleation and growth of new dislocation-poor grains) combined with strain-hardening (generation of new dislocations) to build very complex stress / strain / strain rate / temperature relations. The latter are difficult to summarize into closed-from equations. Alternatively, large sets of experimental curves may be input in a pointwise fashion as the constitutive model; another way consists in coupling plasticity equations with microstructural evolution models (based on dislocation density evolution laws, on recrystallization models, on grain orientation evolution for anisotropy…[21], [22]), although this may be very costly.

## 2.5 Properties prediction

Beyond constitutive modelling and the corresponding measurement techniques, the present challenges lie not so much in flow modelling than in the prediction of final, functional material properties. Indeed, metal forming consists in giving a shape, but most importantly in imparting the part with the desired properties (mechanical but also optical, electromagnetic etc. depending on the application). This goes through microstructure engineering, which is now an integral part of most metal forming modelling. The difficulty is that beyond general microstructure evolution principles, each metal, alloy, composite… develops its own features and must be dealt with using specific physically based material evolution models.

A specific mechanical property is material integrity, i.e. the absence of excessive porosity or crack initiation sites. Damage evaluation is therefore necessary, but the current state of the art is far from satisfactory in this respect [23]. In sheet metal forming, rather than relying on questionable damage function integration, one refers to Forming Limit Diagrams (FLD). They are built from experiments under a set of different triaxiality conditions (pure tension, equibiaxial tension, shear…). Necking strain and fracture strain are measured and reported in a major in-plane strain / minor in-plane strain diagram, giving the boundary between safe and unsafe forming conditions depending on the state of stress (hidden inside the deformation modes). During the FEM computation, the computed strain state is checked and located in this diagram, resulting in a fracture / no fracture diagnosis [24].

There is here ample space for more research, combining different spatial scales, from nano to macro scales.

## 2.6 Practical use in industry

FEM simulation of metal forming sequences has become universal at least in large manufacturing companies. Yet computing time remains an issue for industrial applications. Over the decades, it can be estimated that improvements in algorithms and numerical methods have participated equally with the computer power boom to the general improvement (acceleration of simulations / dealing with more complex problems). Yet the requirements of more complex processes, more local / fine scale microstructural features, multi-physical coupling etc. keep computing time at the forefront of preoccupations. Much hope is put in the development of coupled AI-FEM modelling for faster response times [25].

## 3 Thermal flow

A processing route in metal forming may be summarized in general by a sequence like casting / hot forming / cold forming / assembling, with thermal treatment and surface finishing judiciously positioned in it.

Metal forming most of the time begins with high temperature processes, carried out at ~ 0.7 $T_m$ to 0.8 $T_m$, $T_m$ being the melting temperature in K. The purpose is twofold:

- Benefit from the drop of mechanical resistance around 0.5 $T_m$, in order to reduce drastically the forces, torques and power needed. Indeed, the preceding metal casting processes are necessarily slow because sufficient time must be allowed for solidification and cooling; to obtain a reasonable productivity, the section of the cast product therefore needs to be large. Combined with the high mechanical properties at Room Temperature (RT), this would result in intractable forces and power.

- The casting microstructures are inadequate for most applications, too coarse, with insufficient yield stress, shock and fatigue resistance. Plastic strain above 0.5 $T_m$ allows recrystallization to transform a coarse dendritic microstructure into a fine, equiaxed grain material with high mechanical resistance. This aspect will be developed in Section 5.

On the other hand, hot forming must in some cases be completed by cold forming to improve mechanical properties further (by work-hardening, multiplication of dislocations); for better dimensional control because of smaller tool distortion (reduced thermal dilatation); in view of a better surface finish due to the absence of oxidation. Of course, depending on the alloy and application, the sequence may be terminated after casting (foundry products) or after hot forming (hot forged pieces e.g.), with just machining and/or surface finishing to follow. Heat flow in cold forming will differ from the ones in hot forming, but the difference is more in the orders of magnitude than in its pattern. Figure 2 summarizes the heat flow in the case of a rolling process.

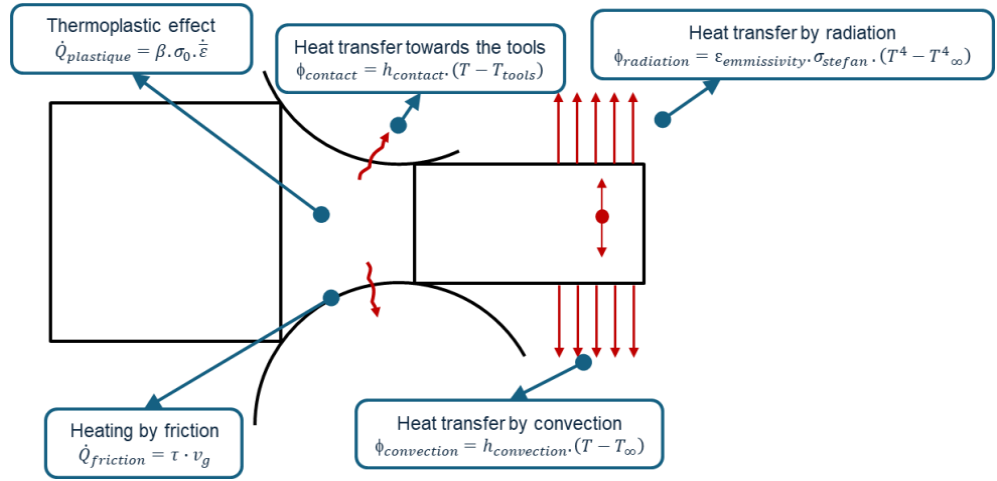

**Fig. 2:** Schematic representation of the heat sources in a metallic material during forming.

In all cases, plastic dissipation occurs inside the metal, whereas friction dissipation occurs at interfaces.

- The power density of plastic heat (per unit volume, W.m$^{-3}$) is given by the product of the equivalent strain rate by the yield stress; β ~ 0.9 to 0.95 is the Taylor-Quinney coefficient (this means that more than 90% of the plastic power is transformed into heat, the rest being stored in the microstructure, e.g. the atomic network distortion by new dislocations, grain boundary energy, stacking fault energy…).

- From this, an estimate of plastic temperature increase can be formed under adiabatic assumption, valid in the centre of a bulky piece:

$$\rho C . \frac{dT}{dt} = \dot{Q}_{pl}(x,y,z) = \beta \sigma_0 . \dot{\bar{\varepsilon}} \quad \rightarrow \quad \delta T = \beta \frac{\sigma_0 . \bar{\varepsilon}}{\rho C} \qquad (1)$$

- In hot forming, with ρC ~ 4-5.10$^6$ J.K$^{-1}$m$^{-3}$, $\sigma_0$ ~ 20 – 200 MPa, temperature increase is generally of the order of 5-40°C for a strain $\bar{\varepsilon} = 1$. In cold forming, $\sigma_0$ ~ 100 – 1000 MPa and ρC ~ 3-4.10$^6$ J.K$^{-1}$m$^{-3}$, plastic heating of 100°C is not uncommon, to the point that temperature may become a limiting factor for speed or reduction, i.e. industrial productivity.

- Frictional heat density (per unit area, W.m$^{-2}$) is the product of the friction (tangential) stress by the sliding (relative) velocity. The heat thus produced at the interface is shared between the tool and the workpiece, according to a Heat Partition Coefficient (HPC). Friction comes mostly from interactions at the roughness scale, the corresponding heat is therefore predominantly *produced* in the material which undergoes microplasticity, i.e. the workpiece. However, the heat will secondarily flow from the warmer to the cooler object. This is generally accounted for through an equivalent convective formulation, $\Phi = h_{contact}(T-T_{tools})$, see Fig. 2, $h_{contact}$ is the Heat Transfer Coefficient (HTC). This redistributes both the frictional and the plastic heat, and the energy present in the hot workpiece in the case of hot forming. This makes this thermal contact problem a quite difficult one, with poorly known HPC and HTCs.

- In cold forming, the sum of bulk plastic and surface frictional heat results in the surface of the workpiece being warmer than the core, by up to a few 10°C; but this is a skin effect and temperature rapidly re-homogenizes after the end of the contact, before the following pass if the process is a multi-pass one. In hot forming, frictional heat is a minor term and the dominant heat flow is from the hot workpiece core (500°C for Al alloys, > 1000°C for steels) to the colder tool, so that the surface is colder than the core. Anyway, a temperature difference always exists in the formed metal, the result may be heterogeneous microstructural evolutions, at least skin versus core. This effect is reinforced by the surface undergoing most of the time more shear than the bulk, due to the subsurface deformation induced by friction; and shear microstructure and crystallographic texture differ from the ones promoted by elongation.

- Workpiece-tool heat transfer can be decreased either by pre-warming the tool, or by cooling superficially the workpiece before the contact to preserve the tools from excessive heat input during the contact, lowering their surface temperature and thermal stress cycles ("skin cooling" in hot strip rolling).

- The workpiece and the tools also exchange with the rest of the environment, by combinations of radiative and convective transfers. The corresponding HTCs can be computed by standard formulae of radiative transfer, free or forced convection with laminar or turbulent fluid flow [26].

- Among these exchanges, tool cooling is the most important. In hot as well as cold rolling, it consists in a permanent water or emulsion spray on the back of the work rolls, as close as possible to the exit of the contact, to be able to extract the heat before it diffuses inward; in hot forging, it is a lubrication and cooling step between two strokes of two successive pieces.

In spite of these precautions, tool warming is a constant concern both in cold and hot forming. There are large superficial gradients (skin effect, boundary layer effect) which result in thermal stresses, whereas bulk heating, although an order of magnitude less, gives differential dilatation which distorts the shape of the tool ("thermal crown" in strip rolling) and may contribute to geometrical defects on the workpiece. The "thickness profile defect" of rolled strip is indeed a combination of roll wear profile, roll thermal crown and roll elastic deformation. Moreover, tool temperature and dilatation generally grow along a mounting, so that the defects themselves are not constant in time, which complicates the corrective settings of the machine tools. If tool cooling is not efficient enough, excessive temperature may soften quenched steel tools by an over-tempering effect and submit them to plastic deformation in the following stages of their lives. Repeated heating may lead to microstructural and phase transformations resulting in brittle surface layers accelerating abrasive wear. These are only a few of the problems heat transfer may cause in metal forming tools. With all this in mind, it is not so surprising that forming tools may sometimes burst…

## 4  Mechanical and thermal contact (tribology)

Friction is a tangential force $F_t$ developed between two *solids* in contact, which resists tangential relative movement (sliding). In case sliding occurs nevertheless, this force dissipates power $\dot{W}_f = F_t.v_g$, where $v_g$ is the relative velocity vector (tangential, "sliding velocity"), see Fig. 2. In metal forming, $F_t$ is applied to the flowing metal, the flow pattern of which it impacts more or less; and the reaction force is applied to the tools, increasing their loading and influencing their damage and wear.

Friction will therefore impact the flow, i.e. the velocity vector field within the plastically deforming metal. This effect may be complex if several tools / interfaces are competing, with possibly opposite effects of friction. Being an interfacial force, friction changes the forming loads (forces and torques). As it dissipates power, it influences temperatures and above all, interface temperatures, with feedback on friction itself (e.g. through degraded lubrication). Finally, it conditions wear mechanisms.

## 4.1 Friction force $F_t$ modelling

In metal forming, it is particularly important to remember that friction is a *threshold phenomenon*. If we assume the Coulomb friction model, it must be written in full:

$$\tau = \mu.\sigma_n \quad if \ \|v_g\| \neq 0 \quad (slip)$$

$$\tau \leq \mu.\sigma_n \quad if \ \|v_g\| = 0 \quad (stick) \tag{2}$$

($\tau$ is the friction stress, $\sigma_n$ the normal stress, $\mu$ the Coulomb friction coefficient). Therefore, in a no-slip area, the friction law does not determine the friction stress $\tau$, just an upper bound of it; only the resolution of the whole boundary problem can give explicitly, everywhere in such a zone, the friction stress; it is likely to be well below $\mu.\sigma_n$.

Moreover, the yield criterion imposes a limit on any shear stress, hence on friction stress:

$$\tau \leq \tau_{max}, \tau_{max} = \sigma_0/\sqrt{3} \quad (von\ Mises\ criterion). \tag{3}$$

Finally, in bulk metal forming, contact pressure is high, often well above the yield stress $\sigma_0$. In such cases, the real contact area is close to 100%, eliminating the fundamental reason for the validity of Coulomb friction, i.e. the quasi-linear growth of the real contact area with applied pressure. As a consequence, another friction law is often used, named the friction factor model or "Tresca" friction law, to enforce the last equation:

$$\tau = \bar{m}.\sigma_0/\sqrt{3} \ (0 \leq \bar{m} \leq 1) \quad if \ \|v_g\| \neq 0 \quad (slip)$$

$$\tau \leq \bar{m}.\sigma_0/\sqrt{3} \ (0 \leq \bar{m} \leq 1) \quad if \ \|v_g\| = 0 \quad (stick). \tag{4}$$

On the contrary, contact pressures in sheet metal forming are generally low (a fraction of $\sigma_0$)[1] and Coulomb's friction law is to be preferred. Non-linear versions $\mu(\sigma_n, T, v_g, [R_q])$ are in general use (T is the temperature, $R_q$ the RMS roughness); it is most of the time found that $d\mu/d\sigma_n < 0$, i.e. friction is not exactly Coulombian, decreasing as pressure increases.

More general forms combining Coulomb at low pressure and Tresca at higher ones are available for complex situations [27], [28]. It must be borne in mind also that, due to often oriented tool roughness patterns, friction may be anisotropic, i.e. (i) $\tau$ depends on the sliding direction and (ii) the friction stress vector direction and the sliding direction may differ [29]. This is rarely taken into account.

## 4.2 Flow patterns and $v_g$

Friction in metal forming can be so high as to block sliding in all or part of the tool-product interface. This is illustrated in Fig. 3 for the plane strain upsetting of a bar, a simple forging operation. This figure also points to a very important feature: the effect of friction depends on the geometry of the plastic deformation zone, such as a surface / volume ratio. The reason is obvious if the relevant equilibrium equation (without inertial or body forces) is considered:

---

[1] One reason is that an out-of-plane normal pressure p applies an in-plane bending stress of the order of p.R/e, where e is the sheet thickness and R the radius of curvature ; in general R >> e → a weak pressure is enough to plastically deform the strip in bending.

$$\frac{d\sigma_{xx}}{dx} + \frac{d\sigma_{xy}}{dy} = 0 \rightarrow \frac{d\sigma_{xx}}{dx} \approx \frac{2\tau}{h} \rightarrow \sigma_{xx} = C + \frac{2x}{h}\tau \rightarrow \overline{\sigma_{yy}} = \overline{\sigma_{xx}} - \frac{2\sigma_0}{\sqrt{3}} = f(\frac{a}{h}\tau), \quad (5)$$

where a superscript bar represents an average in y. Stresses and forces are therefore monotonically increasing functions of $\frac{a}{h}\tau$, the product of an aspect ratio (length a/thickness h) and a friction stress (τ), whatever the friction law used. This is a perfectly general statement for plastic flow. Moreover, the analysis of the state of stress shows that what increases with friction is mainly the hydrostatic pressure p = -1/3 ($\sigma_{xx}+\sigma_{yy}+\sigma_{zz}$), whereas the stress deviator, which is proportional to the strain rate tensor, i.e. depends solely on the flow pattern, is far less impacted.

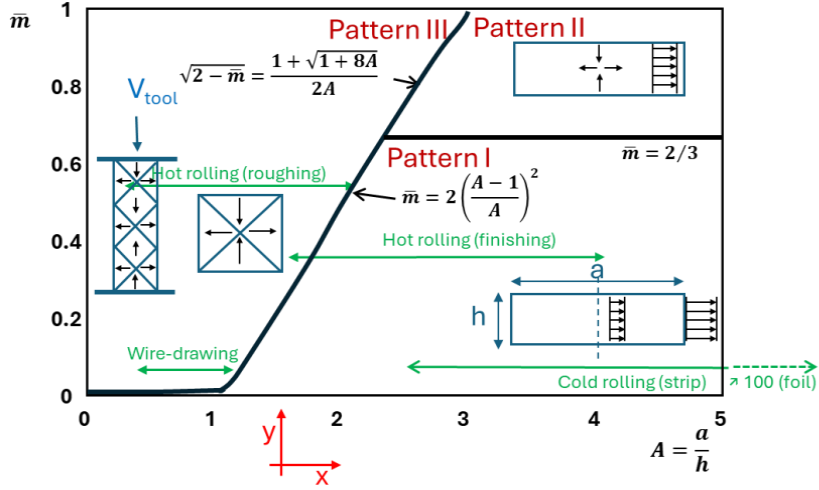

**Fig. 3:** The flow pattern in upsetting (forging) depends on both aspect ratio A = a/h and friction (here, Tresca friction factor $\bar{m}$). Approximate formulae for the total dissipated power $\dot{W}^*$, friction dissipated power $\dot{W}_f$ and plastic dissipation power $\dot{W}_v$ are given below together with the average sliding velocity $\Delta \bar{v}$. The graph is drawn for plane strain but extends qualitatively to axis-symmetry. Hot forging may appear in all sectors I, II or III. Other processes, such as wire-drawing or strip rolling, are positioned both in terms of A = a/h and in terms of friction factor, for comparison.

Pattern I $\quad \dot{W}^*(\bar{m}) = \frac{2\sigma_0}{\sqrt{3}}.2a.V.\left(1 + \bar{m}.\frac{a}{4h}\right); \quad \Delta\bar{v} = \frac{V.a}{4h};$ (6)

$$\dot{W}_f(\bar{m}) = \frac{2\sigma_0}{\sqrt{3}}.2a.V.\bar{m}.\frac{a}{4h}; \quad \dot{W}_v(\bar{m}) = \frac{2\sigma_0}{\sqrt{3}}.2a.V$$

Pattern II $\quad \dot{W}^*(\bar{m}) = \frac{2\sigma_0}{\sqrt{3}}.2a.V.\left(1 + \bar{m}.\frac{a}{4h} + \frac{h}{4a}.\frac{2-3\bar{m}}{2-\bar{m}}\right); \quad \Delta\bar{v} = V.\left(\frac{a}{4h} - \frac{h}{a}.\frac{1}{(2-\bar{m})^2}\right);$ (7)

$$\dot{W}_f(\bar{m}) = \frac{2\sigma_0}{\sqrt{3}}.2a.V.\bar{m}.\frac{a}{4h}\left(1 - \left(\frac{2h}{a}\right)^2.\frac{1}{(2-\bar{m})^2}\right) ;$$

$$\dot{W}_v(\bar{m}) = \frac{2\sigma_0}{\sqrt{3}}.2a.V.\left(1 + \frac{h}{4a}.\left(\frac{4-3\bar{m}+3\bar{m}^2}{(2-\bar{m})^2}\right)\right)$$

Pattern III $\quad \dot{W}^*(\bar{m}) = \frac{2\sigma_0}{\sqrt{3}}.2a.V.\left(\frac{a}{2h} + \frac{h}{2a}\right) = \dot{W}_v(\bar{m}) \quad;\Delta\bar{v} = 0 \quad; \quad \dot{W}_f(\bar{m}) = 0$ (8)

- Large aspect ratio A (flat bar) and moderate friction lead to a quasi-homogeneous strain rate flow, whereby $v_y$ is a linear function of y and independent of x, whereas $v_x$ is independent of y and linear in x (pattern I). No sharp localization nor instability is expected.
- If the aspect ratio is small, strain localizes more and more into diagonals of the cross section, a pattern clearly visible in practice on the hot metal as the "blacksmith's cross"; for very low aspect ratios, sub-diagonals form a number of rigid blocks, all the strain rate is localized. This is pattern III. A plastic instability may occur as a localization shear band at 45° which will result in fracture.

- For intermediate aspect ratio and very large friction ($\bar{m} > 2/3$), a mixed pattern II occurs with a central rigid-block zone surrounded by two quasi-homogeneous strain rate zones.

Note that for A = 1 and $\bar{m} = 0$, two completely different patterns co-exist (I and III): remember that in plasticity, multiple solutions (several flow patterns) are possible; they must just dissipate the same power, which is the case for A = 1 and $\bar{m} = 0$, as shown by the corresponding equations.

### 4.3 Senses of variation and flow pattern

It can be shown mathematically that the total dissipated power (plasticity + friction) is a continuous, monotonically increasing function of friction, moreover with a downward concavity (i.e. the growth saturates at high friction). The *bulk plastic* dissipation $\dot{W}_v$ also increases (as friction triggers more and more *shear*), with possible jumps when the flow pattern changes.

The average slip velocity $\Delta \bar{v}$ is systematically a decreasing function of friction, which is logical since friction resists slip. Hence, no particular variation can be associated with the frictional power since it multiplies an increasing function (friction stress) by a decreasing one (slip velocity).

Hence if friction increases strongly (such as in hot forging compared with lubricated cold forging), flow pattern may change.

These flow patterns are jointly governed by friction and aspect ratios but in turn, they have consequences. Pattern III is characterized by $v_g = 0$, i.e. no frictional energy dissipation, so that power, stresses, forces and torques are independent of friction (see Eq. (8)). Note that the total dissipation is necessarily an increasing function of friction. Hence, as friction increases, the total amount of heat released increases, but the amount released at the interface shows a decreasing trend, whereas heat in the subsurface and bulk increases.

Note that pattern III points to the importance of writing correctly, i.e. with a threshold effect, the friction law, as in Eqs. (2) and (4) above.

Similarly, referring to Preston-Archard wear law [30] : $\dot{m} = K_w . S . \frac{\sigma_n v_g}{Hv}$, $v_g = 0$ cancels abrasive wear ($\dot{m}$ is the mass loss per unit time, S the contact surface and Hv the Vickers hardness). However, this is not the end of tool wear, with other mechanisms such as mechanical and, in hot forming, thermal fatigue, or corrosion / oxidation.

But a consequence is that in metal forming, tool wear is not necessarily a monotonically increasing function of friction, as intuitively guessed: $\sigma_n$ is an increasing function, $v_g$ a necessarily decreasing function of friction: their product may be increasing, decreasing or non-monotonic depending on the specifics of the process.

### 4.4 Effect of friction on different forming processes

- In drawing processes (wire, strip, bar or section drawing), the forming tool (the die) is fixed: the forming power is applied by another, spooling tool (capstan). Therefore, die friction always opposes the forming movement, so that the drawing force is necessarily an increasing function of friction. A good approximation is (wire-drawing from radius $R_e$ to radius $R_s$, die semi-angle $\alpha$):

$$F = \pi . R_s^2 \, \sigma_o . \bar{\varepsilon} . \left( \psi + \bar{m} \frac{\cot \alpha}{\sqrt{3}} \right)$$

$$\sigma_{xx,s} = \sigma_o . \bar{\varepsilon} . \left( \psi + \bar{m} \frac{\cot \alpha}{\sqrt{3}} \right), \qquad (9)$$

where F and $\sigma_{xx,s}$ are the drawing force and stress respectively and $\bar{\varepsilon}= \mathrm{Ln}\left(\left(\frac{R_e}{R_s}\right)^2\right)$ the plastic strain. Compare with Eq. (5), with cot ($\alpha$) playing here the role of the aspect ratio a/h multiplying $\bar{m}$. $1 < \psi(\alpha,\bar{\varepsilon}) < 1.2$, "the redundant work factor", is a shear effect correction to a model which does not involve shear by design.

As the drawing stress is limited to ~0.3 – 0.4 $\sigma_0$ for fear of wire damage or break in pure tension, this puts a limit on the plastic strain $\bar{\varepsilon}$ or the section reduction $1 - R_s^2/R_e^2$. This is why wire-drawing is always a multi-pass operation, with section reduction smaller than 30%. Moreover, as $\sigma_{xx,s}$ grows with friction, the limit becomes more and more stringent as friction increases.
Sliding is imposed, close to the drawing speed (i.e. large). Yet the normal stress decreases slightly as friction gets higher (this is due to the yield criterion: when $\sigma_{xx}$ is larger, $\sigma_{rr}$ must decrease). However, the addition of a tangential component increases tool surface damage. This is why it is systematically found that high friction, with high amount of metal / tool microcontacts, accelerates die wear.

Finally, due to the large sliding speed, the frictional heat dissipation can be a major contribution to interface heating, to the point that the wire surface may reach 200°C of even 300°C for hard steel wire. In certain cases (steel cord) it may trigger Dynamic Strain Aging and lead to wire break or at least, inadequate final microstructure and mechanical properties.
Die friction thus has only detrimental effects, this is the reason why wire-drawing is always lubricated; friction should be reduced as much as possible, under the condition that the finished wire surface state remains within specifications. Note that on the contrary, friction is necessary on the capstan to entrain the wire. N turns of wire are wrapped on the capstan and the "capstan effect" ensures entrainment (capstan friction force proportional to $\exp(\mu.\theta)$, $\theta = 2N\pi$ being the wrapping angle).

- In rolling processes, the forming tools and the power providing tools are the same: the rolls. Coilers / interstand devices apply strip or bar tensions to guide it straight and to decrease the normal load, but are not the primary sources of the forming power. Moreover, the entrainment movement, a rotation, is tangential to the rolled strip: friction is the only driving force (the normal stress, on the contrary, tends to push the rolled product back). As such, friction is indispensable to rolling. But too high friction will anyway cause roll load and torque increase (increased power consumption, roll elastic deformation leading to profile and flatness defects), more heat dissipation hence excessive roll and product temperature, and roll wear. An adequate working interval must be determined for the friction coefficient and different technologies (lubrication in the first row) must be applied to maintain it in this interval, in spite of possibly varying conditions (roll wear, accelerations and transients etc.).
  More precisely, the non-skidding condition (minimum friction to be able to entrain the strip) is given by:

$$\mu_{min} = 1/2.\tan(\alpha_{bite}) \approx 1/2.\sqrt{\Delta e/R} \approx L/2R, \qquad (10)$$

where $\Delta e$ is the thickness reduction, R the roll radius and L the roll bite length. $\mu < \mu_{min}$ leads to *skidding*, with rolling becoming unstable or even impossible. $\mu_{min} \sim 0.2$ for thick products as in hot rolling ($\Delta e$ is large), so that care must be taken with lubrication. In upstream passes of steel hot rolling (e ~ 100 to 200 mm), lubrication is not applied at all and rolls may on the contrary be artificially roughened to ensure entrainment in spite of the large bite angle $\alpha_{bite}$. Anyway, friction is allowed to be large there because, due to the small aspect ratio L/e (<-> A = a/h in Fig. 3), pattern III prevails and friction effect on stresses and forces is small. On the contrary, for the thin formats of cold rolling (pattern I), $\Delta e$ is small and so is $\mu_{min}$, so that a good lubrication may be applied; fortunately, especially for the thinnest products (thin strip, foil).

- Can making is an interesting, specific case. Here, there are two different forming tools, the die which forms the external surface and the punch which calibrates the inner one. As in wire-drawing, the die is fixed, friction is a resisting force; the punch is mobile and it is also the source of the forming power. The can wall slides forward against the die, but backward against the punch due to the elongation of the can: friction stress direction is opposite (entrainment by friction on the punch). Of course, there is dissipation on both interfaces, so that the punch force (or forming power) is an increasing function of both die and punch friction. But looking at the can wall, punch friction is subtracted from the die friction since they have opposite signs: *high enough* punch friction (without excess heating !) results in decreased tension stress in the wall, i.e. a *smaller* risk of break. One may therefore play around with the tool surface states, or differential lubrication, to control separately die friction (very smooth die, as always in drawing processes) and punch friction (may be rougher).

This is a good example to illustrate the following important concept: different levels of friction may be desired or required on different interfaces for optimal working. Similarly, different levels of friction may be optimal at different stages of a given process, as we have seen above about rolling. Friction is not only time- and space-dependent because of the uncontrollable complexity of the contact conditions, but it is often *required* time- and space-dependent. The difficulty is: how to control different friction at different places at different times? This remains an open question for technologists, but some applications are already at work [31].

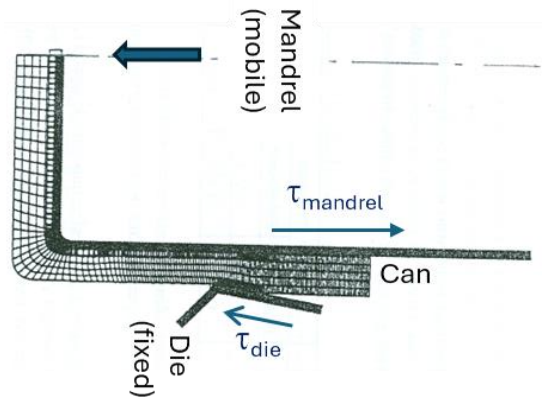

**Fig. 4:** In can ironing, friction felt by mandrel ($\tau_{mandrel}$) and die ($\tau_{die}$) are opposite and therefore play antagonistic roles.

### 4.5 Lubrication

Lubrication is the way to ensure a proper level of friction, specific to each different forming process. Beyond the low friction coefficient, or adequate friction coefficient interval, it must also control the other consequences of friction: heat and wear (see Section 4.6).

The relationship between lubrication and friction is illustrated by Stribeck's curve in Fig. 5. At low speed, a very small amount of lubricant is entrained by the viscous, hydrodynamic effect (see Reynolds' equation below). In fact, no continuous lubricant film may form, lubrication is mainly ensured by polar additives physically or chemically adsorbed on the surfaces: this is the *boundary lubrication regime*. Friction is rather high and, if additives are not efficient enough or if increased temperature prevents their action (i.e. causes their desorption), seizure and galling may occur and friction may increase without bounds.

At the other end, high speed, the viscous effect forms a thick, continuous film separating completely the rough surfaces. This is the *hydrodynamic regime* (or Elasto-HydroDynamic, EHD, if the solid are significantly deformed elastically; or Plasto-HydroDynamic, PHD, if one of them is deformed plastically as in metal forming). Wear is absent (except during transients), friction is low to very low ($\mu = 0.02 – 0.08$ typically), although it increases at the highest speed because the shear stress increases with the shear rate. A drawback of this regime is that just as on a free surface, roughness may grow during plastic deformation, by different mechanisms (plastic instability at large scales, micro-shear bands at grain scale).

In between lies the mixed lubrication regime which borrows its characteristics from the other two: moderate friction and wear, controlled by both additive chemistry and base oil viscosity; some adhesive transfer may occur without degenerating too much, hopefully. Surface roughness is well controlled.

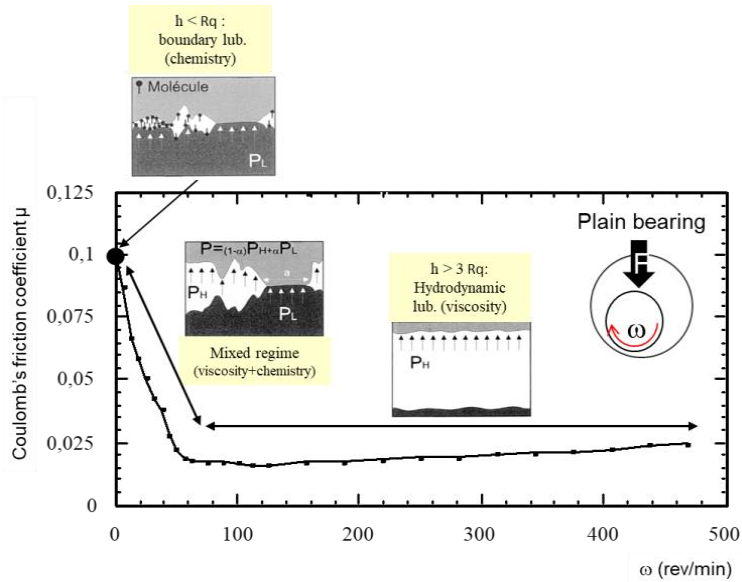

**Fig. 5:** Stribeck's curve in lubrication: effect of velocity on friction coefficient. Example of a steel shaft (Φ 2 mm) in a brass bearing (Φ 2.1 mm), lubricated by a silicone oil η=0.3 Pa.s. Inserts show schematically the change in *lubrication regime*. After [32] .

It must be noted that viscosity, the cardinal property in lubrication, is by no means a simple property of fluid. Its dependence on temperature, but also pressure and shear rate, is crucial and very difficult to measure completely: data are often missing and this remains an open field for research.

Yet a quantitative approach is possible. The above-described behavior can be understood using Reynolds' equation. It is based on a simplified solution of Stokes' equations in thin films using the Hydrodynamic Lubrication Hypotheses (HLH), combined with mass conservation equation integrated through the film thickness h (in the z-direction):

$$\frac{\partial}{\partial x}\left(\frac{\rho h^3}{12\eta}\frac{\partial p}{\partial x}\right) + \frac{\partial}{\partial y}\left(\frac{\rho h^3}{12\eta}\frac{\partial p}{\partial y}\right) = \frac{\partial}{\partial x}\left((V_{x,1}+V_{x,2})\frac{\rho h}{2}\right) + \frac{\partial}{\partial y}\left((V_{y,1}+V_{y,2})\frac{\rho h}{2}\right) + \frac{\partial \rho h}{\partial t}. \quad (11)$$

It relates the fluid pressure field p(x,y) to the opening field (thickness h(x,y)) and the solids superficial velocity vectors **V**$_1$ and **V**$_2$. Dynamic viscosity η(T,p) and volumic mass ρ(T,p) are the two fluid properties involved.

Assuming Plane Strain and steady state for a simpler form:

$$\frac{\partial}{\partial x}\left(\frac{h^3}{12\eta}\frac{\partial p}{\partial x}\right) = \frac{\partial}{\partial x}\left((V_1+V_2)\frac{h}{2}\right). \quad (12)$$

It can be integrated easily with adequate boundary conditions to give the lubricant film thickness $h_c$ in e.g. wire-drawing ($V_1 = V_{wire}$, $V_2 = V_{die} = 0$), assuming pure hydrodynamic regime between smooth surfaces, with the viscosity described by η = η$_0$.exp(γp) (Barus' model of piezoviscosity):

$$h_c = \frac{3\eta_0 . \gamma . V_{wire} . \cot \alpha}{1 - exp(-\gamma \sigma_0)} \qquad (13)$$

α is the die semi-angle as above (generally 6°). The denominator is generally close to 1 since $\gamma \sigma_0 \sim$ a few units, hence

$$h_c \approx 3\eta_0 . \gamma . V_{wire} . \cot \alpha. \qquad (14)$$

Estimates can be built in the same way for most of the usual forming processes [33]-[36]. Numerical application shows that thickness between 100 nm and 10 μm may be expected depending on the type of lubricant and the process. The target depends on the process. Taking the same examples:

- In wire drawing, one tries to minimize friction and die wear. Thick hydrodynamic films are therefore welcome. This is why soaps are used as lubricants for steel and copper wire-drawing: as interface temperature grows, they melt through a series of highly viscous mesomorphic phases (smectic liquid crystals). However other kinds of lubricants may be used, in particular when cleanliness is at stake: light oils for aluminum to avoid stains after annealing, emulsions or dispersions of fats in brass-coated steel-cord "immersed" wire-drawing for a better temperature control.
- In strip rolling, hydrodynamic regime is quasi-forbidden as significant friction is *needed*. The mixed lubrication regime, even close to the boundary regime, is required. Carbon steel cold rolling is almost exclusively lubricated with oil-in-water (O/W) emulsions, also to ensure a complement of cooling (each stand is equipped with specific strip and roll cooling sprays). Note that emulsion lubrication involves preferential entrainment of oil by wetting of the roll and strip surfaces, by a kind of pressure-driven inversion of the emulsion, so that only (or mainly) oil goes through the roll bite; if water penetrates, performance degrades. The low amount of oil (1 to 5% in water) allows controlling the film thickness through the starvation phenomenon [33]-[38], playing around with the emulsion stability, the droplet size, the concentration…Such a lubrication involves oil viscosity engineering, emulsifier physical chemistry, boundary additive chemistry, making it a particularly intricate formulation problem. By contrast, stainless steel cold rolling lubrication involves mainly low-viscosity oils ("kerozens", kinematic viscosity 8-12 cSt at 40°C) with polar and extreme pressure additives. The purpose is to obtain a very bright surface (very low roughness) which can be obtained only with very smooth ground rolls and, of course, boundary lubrication regime. Only those products from which bright surface aspect is not required can be lubricated with emulsions.
Similarly, light alloys are lubricated with kerozens (typically 2-4 cSt at 40°C) with polar additives (fatty alcohols and acids) but the reason is different: to avoid stains at annealing, again.
More generally, lubricant composition is necessarily alloy-dependent to a large extent, because the surface reactivities of the metals and oxides vary greatly: between acid and basic oxides, the components that must react, and those which shouldn't react, are necessarily different.
- Another point which requires careful control of the lubricating system is that lubricants evolve with time, for better or for worse. One of the reasons is the progressive pollution of the lubricant bath (a steel rolling oil tank volume may be > 100 m$^3$, by the way) by metal fines, i.e. wear particles, unavoidable due to the very strong interaction between tool and formed metal. Their reaction with e.g. free fatty acids leads to soap formation which may improve lubrication – or not. Additives are consumed by reaction with solid surfaces, so that their concentration tends to decrease, which must be compensated regularly (lubricant system maintenance, replenishment on the basis of chemical measurements). Emulsion oil droplet size may evolve due to repeated shearing in the nozzles, etc.
- Lubricants are generally recirculated, with filtration on the circuit for cleanliness purposes. Pumping lubricant from the tank to the spraying nozzles is a significant energy cost. Disposal at life end is another one, in particular for emulsions.

Lubrication / cooling interferences are a major problem is metal forming. Limitations of speeds, reductions… in other words, productivity, are fixed by the temperature reached at the interface, when it exceeds the maximum working temperature of the lubricant (e.g. desorption temperature of oiliness additives), giving way to adhesive wear (transition from mild to severe wear). This is of course not the only source of limitation: we have seen another cause above, with hard steel wire-drawing speed being limited by wire plastic and frictional heat leading to microstructural degradation. Different operations and different alloys have their specificities. Dissipated heat must therefore be extracted in particular from the hot forming tools to avoid damage, superficial crack (fire-cracking). In cold forming, the problem is more dilatation and imprecise geometry. In some processes (strip rolling typically), both cooling and lubrication are so important that they are performed separately, with different sprays for cooling (on the back of the work roll to extract heat as soon as it is produced) and for lubrication (in front, near the bite entry point). Water-based lubricants, namely emulsions (or more rarely, dispersions) of oil in water (O/W emulsions) are preferred whenever a high cooling power is needed.

## 4.6 Tool wear / adhesive transfer

In general, a material A may scratch / abrade a material B if its hardness $Hv_A \geq 1.2\ Hv_B$ [39]. Tool steels or ceramic tools are much harder than the material they form. How comes tools may undergo abrasive wear? The reason is that the formed metal may be macroscopically soft, while containing hard second phase particles which, arriving at the surface due to surface area growth, may scratch the tool surface. To avoid this, tools may be reinforced by e.g. a large proportion of hard carbides (VC, WC in high speed steels) or by coatings such as hard chromium. Note that the latter should be abandoned because its deposition involves toxic $Cr^{VI}$ salts. Search for alternatives has been very active since ~30 years, without credible results; this is a hot topic in cold metal forming tool engineering.

Other wear mechanisms exist:

- oxidation (a form of corrosion) for hot forming tools, with peeling-off of the oxide layer when it becomes too thick;
- fatigue: mechanical fatigue results after very long series (thousands of pieces at high frequency in cold forging e.g.). Thermal fatigue in hot forming is enhanced by high temperature cycles tempering the tool metal. Oligocyclic fatigue of roughing mill rolls in the steel industry is well known: thermal stress are very high in the sub-surface due to the fast-varying, very high temperature gradient (thermal skin effect [40]); they may cyclically reach the plastic yield stress and damage follows by ratchetting. It is often associated with "fire-cracking", a network of shallow cracks which accelerate abrasive wear.

On top of this, forging tools e.g. may undergo plastic deformation which leads to their end of life. This may be because of too high contact temperatures in hot forging (over-tempering the quenched tool metal), or due to extremely high stress on protruding too small radii.

Another very important degradation mechanism is adhesion of the formed metal onto the tool surface. It completely changes the tool properties, in the first row, friction. In hot rolling of light alloys, in case of poor lubrication, aluminum plates may even macroscopically adhere to the roll and bring about serious damage to the equipment. Less harmful, microscale adhesion is quasi universal. It creates undesired roughness on tools which are supposed to be smooth, increasing friction and degrading the surface quality of products [41], [42].

## 5 Microstructural evolution

## 5.1 Introduction (what is a microstructure?)

The microstructure of a metallic alloy is somehow its personality, it defines its behavior (mechanical, thermal, electrical…) and is inherited from its chemical composition and all its thermo-mechanical history (processing route and in-use). In polycrystalline materials, the microstructure can be defined as the arrangement and distribution of several constituents at different scales, such as crystallographic phases, grains and crystalline defects as illustrated in Fig. 6. Each of those entities can be defined with a set of parameters that evolve continuously during the material forming because of plastic deformation and/or temperature. Hence, the definition of a processing route depends on the initial and desired final microstructure. Even more, the aim of metal forming is not only to shape geometrically industrial pieces, but also to control and obtain some specific microstructures, in order to have some control of the final in-use properties. This microstructure control has become increasingly important in the last decades since our requirements in terms of in-use properties are highly increasing.

In a crystalline material the microstructure can be defined as the crystalline structure and the defects. The presence of those defects defines the unstable nature of polycrystalline materials which makes its microstructure evolve under high temperature or mechanical stresses (thermo-mechanical conditions). Here are some of the most studied lattice defects because of their high influence on alloy properties. Most of them are illustrated in Fig. 6.

- 0D point defects (vacancies) will have a small influence on mechanical properties. However, this is the main defect that influences diffusion and thus influences diffusive phase transformation kinetics. Vacancies concentration is highly influenced by temperature, however, in some configurations, dislocation interactions can generate some vacancies [43].

- 1D line defects (dislocations) are mainly induced by plastic deformation, it is their behavior (Burgers' vector, slip systems, glide, cross slip, climb, Peierls stress, partial dislocations…) which defines the mechanical behavior of the material. Additionally, the stored energy (mainly elastic) induced by dislocation is the driving force for recovery and recrystallization. Dislocations are also known to highly influence phase transformation through pipe diffusion and heterogenous nucleation [44].

- 2D surface defects are the boundaries between regions with different crystal orientation (grains) or with different crystal structure (phases). They play a major role in the mechanical behavior of a material (strength, fatigue, ductility…). Being able to control the grain size in a metallic alloy is crucial for industrial applications. This grain size control can only be done with a control of recrystallisation kinetics. Stacking faults should be also mentioned as surface defects. In some alloys, such as Face-Centered Cubic aluminum (or some Hexagonal Close-Packed alloys), some dislocations can split into partial dislocations whose Burgers vector is not a lattice vector. Hence, the slip of partial dislocations leaves behind an imperfect crystal containing a stacking fault.

- 3D volume defects (voids and micro voids) are generally inherited from solidification or damage. Industrially they are generally detected with non-destructive ultrasound measurements. They significantly degrade the mechanical properties and parts with porosity are generally rejected. Void closure and healing can take place during forming processes with negative triaxiality (compressive state of stress) such as forging [45]. This phenomenon highly depends on the deformation thermo-mechanical conditions (stress state, strain, temperature, strain rate and path), on the voids characteristics (size, morphology, distribution and content) and on material properties (flow stress and diffusion coefficients). Some micro-voids can be created because of damage during forming with positive triaxiality (sheet metal stamping for example). Those micro-voids can lead either to failure, or to a degradation of mechanical properties.

The evolution of a microstructure is mainly defined by the evolution of those defects, in addition to the crystalline structure of the present phases and their crystal orientation (texture and micro-texture). Those evolutions are induced by the activation of different physical mechanisms, operating on several scales. The control of those mechanisms and their kinetics during material forming can allow a control of the microstructure and of the in-use properties.

In the following, the most important and generic of those physical mechanisms that can be activated during metal forming are presented: plasticity, recovery and recrystallization, phase transformation and their interactions. This choice of presentation was made to keep a generic and physically based (and not empirical) view of physical metallurgy in the context of materials forming. Otherwise, it would have been necessary to present the specifics of different alloys and forming process which would have given a long list. The main inconvenient is that the reader will not be familiarized with some classical transformation routes of industrial use (for some steels or aluminum alloys for example) but will have the basics to understand the underlying physics behind them.

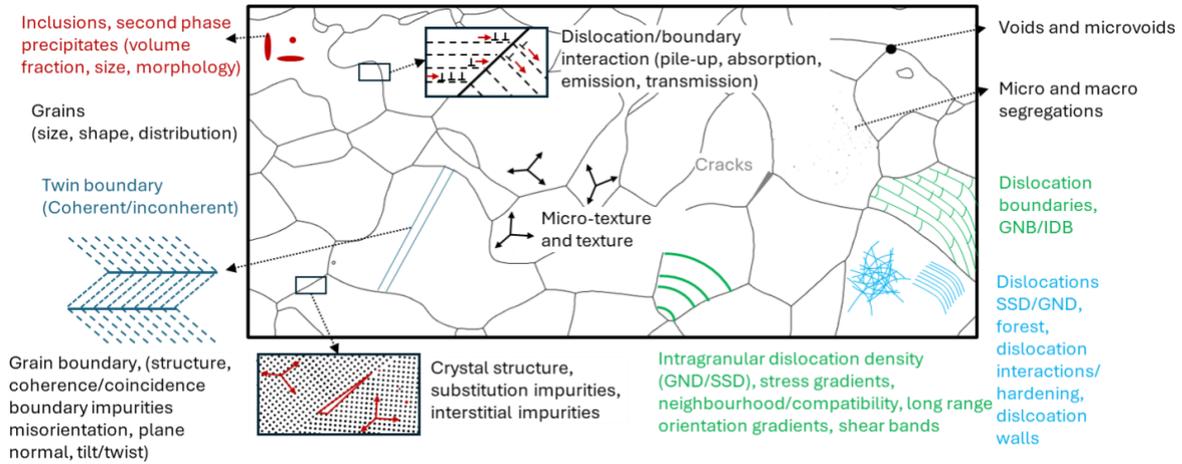

**Fig. 6:** Schematic representation of some entities that constitute a microstructure (GND: Geometrically Necessary Dislocations, SSD: Statistically Stored dislocations, GNB: Geometrically Necessary Boundary, IDB: Incidental Dislocation Boundary).

### 5.2 Physical mechanisms of plasticity

The continuum mechanics approach allows studying the material flow and the mechanical behavior at macroscopic scales. In this approach the material is considered as continuous and homogeneous. Even though some models do use microstructural quantities such as dislocation density, the microstructure is not considered in those models that are mainly phenomenological. In this section, plasticity is presented as a physical mechanism operating at different microstructural scales, inducing changes and leaving traces that will affect any further microstructure evolutions.

#### 5.2.1  *Strain $\varepsilon$ vs dislocation density $\rho$*

In crystalline materials, plastic deformation occurs by the slip of atomic planes. It does not take place at one time but through dislocations. This slip takes place on planes with the highest atomic density in the directions with the highest density. The reason for this is that the closer the atoms are, the smaller is the necessary shear stress to activate slip, i.e. smallest Burgers vector which represents the atomic displacement produced during the slip of a dislocation. This combination of slip plane and slip direction is called slip system which depends on the crystal structure of the material. The total slip of all dislocations induces the macroscopic scale strain as illustrated in this equation [43]:

$$\varepsilon = b\rho_m \bar{x} \qquad (15)$$

$$\dot{\varepsilon} = b\rho_m \bar{v} \ , \qquad (16)$$

where $\rho_m$ is the density of mobile dislocations, b the magnitude of the Burgers vector, $\bar{x}$ the average distance of dislocation slip and $\bar{v}$ the average dislocation velocity. "The value to be attributed to [$\bar{x}$] and its variation with strain are responsible for much of the uncertainty in current theories of work hardening" [46].

Those equations demonstrate that it is impossible to directly link the total dislocation density (and not only the mobile ones) with the plastic strain. This constitutes the main difficulty in establishing physically based constitutive laws for plasticity at the macroscopic scale (a detailed review of physically based work hardening models can be found in Kocks and Mecking [47]). Hence, a clear separation between the two quantities, strain and dislocation density, should be considered. Strain ε, is a history variable indicating the amount of flow that was applied, it is an indication on the amount of slip that occurred within the microstructure. Dislocation density ρ, can be considered as an internal state variable associated with stored energy induced by the presence of dislocations. Unlike strain, dislocation density does not necessarily indicate the slip activity, it rather indicates the dislocation storage.

*5.2.2    Schmid factor and crystal plasticity*

As already mentioned, slip occurs on a slip system. Therefore, the slip behavior highly depends on the local crystal orientation, since the amount of stress resolved on the slip plane and in the slip direction depends on the alignment of the slip system and the applied load. The shear stress resolved on the slip plane in the slip direction, can be expressed:

$$\tau = \frac{F}{A} \cdot cos\phi \cdot cos\lambda , \qquad (17)$$

where $\phi$ is the angle between the load and the normal to the slip plane and $\lambda$ the angle between the load and the slip direction. $cos\phi \cdot cos\lambda$ is known as the Schmid factor.

The critical resolved shear stress, $\tau_c$, is the shear stress necessary to activate a slip system and to make a dislocation slip. The increase of the macroscopic flow stress observed during plastic deformation, i.e. strain or work hardening, is induced by the interactions of dislocations and not by an increase of $\tau_c$ which is a material property.

One of the most important aspects of Eq. (17), is that it makes it clear that the resolved shear stress does not depend only on the applied load, but also on the local crystal orientation. Hence, plasticity is anisotropic and non-homogenous in a polycrystalline material. At the macroscopic scale, in order to obtain an isotropic homogenous response, grains size must be small compared to the sample size and the crystal orientation of the grains should be random with no favored crystal orientation (no crystallographic texture). Additionally, when a slip system is activated and dislocations are generated, the local crystal orientation evolves. Hence plastic deformation can induce the formation of crystallographic texture towards more stable orientations which depends on the microstructure of the initial state and on the deformation conditions (strain rate, strain path…). Hence, an important number of studies can be found on the texture development of some specific materials during some specific deformation mode [48], [49]. Development of crystal plasticity models to predict the mechanical behavior and the evolution of crystallographic orientation is an active research topic [50] where the increase of Statistically Stored Dislocations (SSD) and Geometrically Necessary Dislocations (GND) is described on the basis of the macroscopic strain and local activated slip systems.

*5.2.3    Deformed state and spatial distribution of dislocations*

The presence of dislocations induced by plastic deformation increases the stored energy of the material. Hence, recovery and recrystallization may take place when a temperature increase allows for dislocation and boundary mobility. Additionally, it is known that dislocations have an important effect on phase transformations [44]. It is therefore necessary to have a good quantitative description of the deformed state in terms of dislocations density and spatial distribution. A general schematic description of dislocation distribution is illustrated in Fig. 7. The formation of dislocations sub-boundaries depends on the material and the ability of dislocations for 3D slip, mainly through cross slip. This latter mechanism can take place in FCC materials with medium to high stacking fault energy (no or small stacking faults), in BCC and in some HCP materials since slip is possible on several slip systems with the same Burgers vector. Hence, the number of dislocations within the sub-boundaries (Low Angle Boundaries LAB in opposition to High Angle Grain Boundaries HAGB) depends on the material, for some materials no sub-boundaries are observed. It should be mentioned that this spatial distribution of dislocation is energetically favorable since it reduces that elastic strain field around a given dislocation by compensation with other dislocation (tension vs compression). An important amount of scientific work to develop a universal theory of sub-structure formation in deformed metallic alloys can be found. However, *"This absence of a statistically based non-equilibrium theory is an important limitation in our ability to describe the collective behavior of dislocations that governs almost all aspects of the mechanical properties of materials"* [50]. In Figure 7a, a schematic representation illustrates the spatial distribution of dislocations within sub-structures. This general description does not directly apply for all alloys but should be considered as a general tendency where in some cases, Fig. 7f, no sub-boundaries are formed and all dislocations are forest type dislocation like those represented inside the substructure as illustrated in Fig. 7a. Deformed purer copper microstructures are presented in Fig. 7b and e, the comparison shows that the increase of strain leads to a very well-defined formation of sub-boundaries, that become HAGB. The formation of sub-structure in Al-Cu-Mg alloy is presented in Fig. 7c to demonstrate that sub-boundaries can be present in alloys and not only in pure metals, even though their formation is less clear and sub-boundaries are thicker. The comparison between Fig. 7d and f demonstrates that, in the same material, the spatial distribution of dislocations is dependent on the crystal orientation; some orientations are favorable for the formation of sub-structures and other are not. This difference induces important variabilities in mechanical behavior and further recrystallization, which may have an important effect on the material forming conditions. Such heterogeneities need to be controlled, and thermo-mechanical conditions need to be designed to account for them. As illustrated in Fig. 7e, severe plastic deformation can be used to generate polycrystals with very small grain size (<1μm) known to have a particular mechanical behavior.

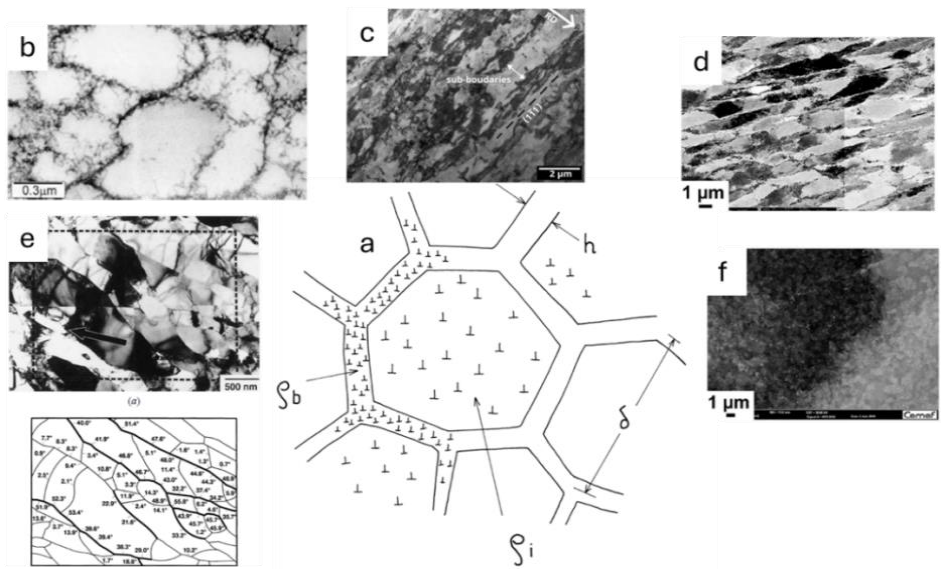

**Fig. 7:** Deformed microstructure at the scale of sub-grains: **a.** Schematic representation [51], **b.** 25% cold-rolled copper [46], **c.** 50% cold-rolled Al-Cu-Mg alloy (AA2024) [52], **d.** pure tantalum deformed by 25% cold compression in a γ-fiber grain (<111> direction parallel to compressions direction) [53], **e.** pure copper deformed by equal channel angular extrusion (ECAE) with 8 passes [54] and **f.** 38% cold-rolled tantalum in a θ-fiber grain (<100> direction parallel to normal direction) [53].

The density and the spatial distribution of dislocations in a deformed metallic alloy depend on several parameters related to the deformation conditions on one side (temperature, strain, strain rate, strain path…) and on the material itself (purity, crystal orientation, phases…). For instance, it was recently demonstrated that the presence of second phase particles does affect the morphology, the orientation and its gradient in Al-Cu-Mg alloy [52]. In this section the deformed state obtained at room temperature was only discussed, hot deformation will be discussed in the following section.

Note: it should be mentioned that dislocation slip is not the only physical mechanism of plasticity. It can be induced by a phase transformation with Transformation Induced Plasticity (TRIP) and by twin formation, TWinning Induced Plasticity (TWIP). Those mechanisms are not discussed in the present paper.

### 5.3 Recovery and recrystallization

At high temperature, the mobility of dislocations and grain boundaries is activated which makes recovery and recrystallization take place, if the amount of stored energy exceeds a given threshold. The term static recrystallization is used when the heat treatment is taking place after the plastic deformation (wire drawing for ex.). The term dynamic recrystallization is used when plastic deformation is done at high temperature (hot forging or hot rolling for ex.), so that all the physical mechanisms are activated simultaneously. After hot deformation, the temperature of industrial pieces stays elevated for several minutes because of thermal inertia (or process organization); if the amount of stored energy within the material is still high enough, recrystallization can still be active. In this case it is defined as Post Dynamic Recrystallization. It can take place either with the appearance of new grains (static recrystallization) or without it, just by the migration of already existing HAGB (meta dynamic recrystallization). A detailed analysis of recrystallization can be found in [46].

Whenever plasticity is combined with high temperature (simultaneously or subsequently), recrystallization will take place. It is the only mechanism (except for the case of severe plastic deformation) that induces the creation of new HAGB which allows the decrease of grain size. Additionally, the process forming routes are generally defined in a way that allows removing the solidification microstructure (with very big grains). Finally, recovery and recrystallization are the only mechanisms that lead to the decrease of dislocation density, which increases the ductility. Hence, recrystallization is of great industrial interest and modelling its kinetics is crucial to define the transformation routes.

#### 5.3.1 *Recovery*

As recovery operates on dislocations, it is activated at lower temperature than recrystallization. Slip and cross slip are activated, redundant dislocations can annihilate (those with opposite Burgers vector) and the others can form sub-boundaries, since this position reduces the energy of each dislocation. Recovery is increasingly active in material in which cross slip can take place. The main consequence of recovery is the decrease of dislocation density, and consequently of stored energy, slowing down subsequent recrystallization. This behavior has been discussed in the literature [46]. However, in materials where cross slip can take place, sub-boundaries can be created during recovery (from line defects to surface defects), and the entangled dislocations on the already existing sub-boundary, can disappear making sub-boundaries mobile without incubation time during subsequent recrystallization. This behavior, which can be observed in the early stages of recovery for low to medium deformed material, was recently reported [55]. Additionally, the activation of recovery may have an influence on the crystal orientation of recrystallized grains [55], [56].

#### 5.3.2 *Recrystallization*

Recrystallization can be defined as the appearance of new grains (new HAGB) that grow. It can take place as a nucleation and growth mechanism (discontinuous recrystallization) or in a progressive way where dislocations form sub-boundaries with increasing disorientation until they become HAGB (continuous recrystallization). The driving force for recrystallization is the stored energy induced by the dislocation density. Stored energy cannot be directly measured but is mainly estimated with two methods. The first one considers the sum of the energies of all dislocations:
$$E_S = \alpha \rho G b^2 , \tag{18}$$

where α is a constant that can be approximated to 0.5 [51] and G is the shear modulus.

The main assumption behind this estimation is that it considers that the elastic strain fields around different dislocations do not interfere, which is particularly false when sub-boundaries are formed. The second method consists of considering that all dislocations form interfaces (sub-boundaries), hence the stored energy is the sum of the energies of all interfaces:

$$E_S = S_v \gamma ,  \qquad (19)$$

where $S_v$ is the total sub-boundary area per unit volume and $\gamma$ is the sub-boundary energy that can be estimated by the Read-Shockley model [57].

The main assumption of this second method is that it neglects the energies of individual dislocations. Hence, it is preferred to combine both methods [58] on the basis of the nature of dislocations (individual forest dislocations vs sub-boundaries dislocations), as illustrated in Fig. 7a.

Dynamic recrystallization is of particular importance when it comes to material forming (hot forging, hot rolling…). Generally, several hot forming steps are necessary to "remove" the solidification coarse microstructure. Unlike static and post dynamic recrystallization where the recrystallization fraction evolution in function of time is studied to model the kinetics, in dynamic recrystallization, the evolution of recrystallized fraction is studied in function of strain. Experimentally, the use of stress-strain curves obtained from hot compression (transient state in general) or hot torsion (steady state in general) are used in addition to microstructural observations. The steady state observed in terms of stress can be directly related to deformation conditions (temperature and strain rate) and grain size [59]. A detailed review of dynamic recrystallization can be found in Ref. [60].

The crystal orientation of recrystallized grains is a subject of great industrial interest, for example in the electric steels field [61]. The recrystallization texture is mainly defined by the orientation of the new grains (oriented nucleation theory) and the relative growth of those new grains (oriented growth theory). Despite an important polarization between the two theories for long time, it is nowadays accepted, thanks to the development of synchrotron capacities, that the orientation of the nuclei is inherited from the deformed state [62] and orientation relationships highly influence boundary mobility [63].

When a microstructure is fully recrystallized, grains keep on growing in the so-called grain growth regime, even though no stored energy is still present. The velocity of this capillarity-driven growth depends on the energy and the mobility of each boundary (which can be defined with 5 parameters) making it particularly difficult to develop high fidelity modelling strategies that account for all the intrinsic physics of the phenomena [64].

Second phase particles highly influence recrystallization, either indirectly by modifying the spatial distribution of dislocations as already discussed [52], or directly by accelerating nucleation on particles (Particle stimulated nucleation) and slowing down grain boundary migration by the Smith-Zenner pinning mechanism. It should be mentioned that the investigation of the fundamentals of recrystallization is still an active research field with several open questions

### 5.4 Phase transformation

As discussed previously, several phases can be present in a metallic alloy. Phase transformation in metals can be divided into two types:

- diffusional transformation requires the migration of atoms (diffusion) for distances higher than the atomic spacings. These phase transformations induce local chemical composition modifications and are generally relatively slow with transformation kinetics highly dependent on the diffusion coefficients of the involved species;

- diffusionless transformations also called displacive (like martensitic transformation in steels for example) involve the synchronous displacement of several atoms for distances smaller than atomic spacing. This transformation is almost instantaneous since it does not involve diffusion.

This section is not dedicated to a detailed presentation of phase transformation but only to some particularities that are directly related to material forming. For a detailed presentation of phase transformations the reader can refer to [65], [66]

In the context of material forming, it is very important to keep control over the phases' volume fraction all along the transformation route. Ideally, the forming is done with the material with the lowest yield strength. This aspect defines some particularities for every class of materials. Here are some of the well-known examples.

- Martensitic steels are mainly formed in the austenitic domain (high temperature) and the martensitic quenching take place when the final geometry is obtained (only very small additional deformations can take place). This transformation route takes advantage of the soft and ductile austenite for forming and the hard and brittle martensite for in-use properties. Some additional heat treatments can be done after quenching, in the austenitic domain, in order to activate carbon diffusion and relax martensite which reduces the hardness of the martensite but also its brittleness.

- Precipitate hardened aluminum alloys (2XXX, 6XXX and 7XXX) are generally formed after a solution heat treatment where all precipitates are dissolved (except Guinier-Preston-Bagaratsky). Hence the material is relatively ductile with reduced hardness. Once the final shape is obtained, a natural and/or artificial ageing is applied to precipitate hardening phases.

- Some Nickel based superalloys (Inconel 718 for example) can have multiple phases within the matrix ($\delta$, $\gamma'$ and $\gamma''$). $\gamma''$ is the main hardening phase and needs to be dissolved for forming. $\delta$ can make a great contribution in limiting the grain size during hot forging thanks to the Smith-Zener pinning mechanism. Hence, some of the transformation steps are done in a temperature domain where $\gamma''$ is dissolved while some $\delta$ phase is still present. Hence, the forging window is between the solvus of those two phases. In alloy design, it is important to consider the forging temperatures to establish the chemical composition of the alloy.

An additional important aspect that defines the limitation of in-use properties in terms of temperature is phase stability. Indeed, any precipitation/dissolution during the use of a metallic piece will highly modify its in-use properties. This is true for volumic phases but can be extended to surfaces and interfaces, at which segregations from the volume as well as reactions with the environment (oxidation or corrosion) may take place either during forming or in-use.

## 5.5 Physical metallurgy in materials forming

The main physical mechanisms that should be considered when studying material forming have now been presented. However, the control of those physical mechanisms and their kinetics for properties control is becoming increasingly difficult. Three main reasons can be listed:

- The requirements in terms of in-use properties are becoming more stringent. This requires on one side the development of new alloys and their corresponding transformation routes. On the other, it pushes towards highly optimized forming processes.

- Some relatively new and more complex processes are developed in order to form more complex pieces in terms of geometry and properties. Flowforming [67] is an incremental, high speed process. Because of those particularities, recrystallization and phase precipitation can take place with highly accelerated kinetics at highly reduced temperature [67]. The interactions between those physical mechanisms are very complex and far from being fully understood and accurately modelled. So, it is still impossible to predict microstructural evolutions and fracture in such cases. Another example is the hot form quench process on aluminum alloys [68] where, unlike conventional forming routes, the material is formed at high temperature with cold tools (forming and quenching simultaneously). This process leads to very complex microstructural evolutions with plasticity, recrystallization and precipitation taking place at the same time, under anisothermal conditions.

- It is expected that the use of recycled materials will increase in the next decades. Hence, forming routes, already optimized for some specific alloys, need to be modified to account for materials with lower formability and unconventional impurities. Additionally, it would be difficult to guarantee a consistent chemical composition for some recycled alloys over the years, which indicates that the forming route should account for such potential evolutions or variabilities.

Those three reasons make it impossible to develop and optimize forming routes empirically as it has been in the past. It is necessary to integrate the metallurgical understanding and modelling at the different microstructural scales upstream of forming process and not only downstream as it has been done till now.

# 6     Conclusion

The present paper is a short introduction to material forming, where general well-established knowledge was presented. The four aspects that constitute the pillars of metal forming have been presented: material flow, thermal flow, mechanical and thermal contact, and microstructural evolutions. This choice of presentation allowed metal forming to be examined from distinct disciplinary perspectives, providing a more structured and comprehensive understanding of the underlying phenomena. For each of those four aspects, in addition to presenting the established and accepted knowledge, current research topics and open scientific questions have also been briefly introduced.

Finaly, the development of forming processes was carried out with very little connection between these four aspects. Today, with increasingly important requirements (higher-performing properties, use of recycled materials, and increasingly complex processes), it is necessary to introduce these four aspects and find a way to integrate them all upstream in the development and design of new forming processes.

# References
[1]  G. Dieter, *Mechanical metallurgy*. Mcgraw-Hill Book Company, 1988.
[2]  J.-L. Battaglia, *Heat transfer in materials forming processes*. John Wiley & Sons, 2008.
[3]  S. R. Schmid, *Schey's Tribology in Metalworking—Friction, Lubrication, and Wear*. ASM International, 2023.
[4]  *Physical Metallurgy*. Elsevier, 2014. doi: 10.1016/C2010-0-65716-6.
[5]  P. Montmitonnet, « Laminage - Objectifs et enjeux de la modélisation », *Mise en forme des métaux et fonderie*, juin 2016, doi: 10.51257/a-v2-m3065.
[6]  H. Tresca, « Mémoire sur l'écoulement des corps solides soumis à de fortes pressions », *Comptes Rendus de l'Académie des Sciences*, vol. 59, p. 754, 1868.
[7]  R. Wagoner, J.-L. Chenot, et W. Knight, « Metal Forming Analysis », *Applied Mechanics Reviews*, vol. 55, nº 3, p. B52-B53, mai 2002, doi: 10.1115/1.1470681.
[8]  R. Hill, *The Mathematical Theory of Plasticity*, Clarendon Press, Oxford. 1950.
[9]  F. Barlat, S.-Y. Yoon, S.-Y. Lee, M.-S. Wi, et J.-H. Kim, « Distortional plasticity framework with application to advanced high strength steel », *International Journal of Solids and Structures*, vol. 202, p. 947-962, oct. 2020, doi: 10.1016/j.ijsolstr.2020.05.014.
[10] A. N. Bramley et F. H. Osman, « The Upper Bound Method », in *Numerical Modelling of Material Deformation Processes*, P. Hartley, I. Pillinger, et C. Sturgess, Éd., London: Springer London, 1992, p. 114-130. doi: 10.1007/978-1-4471-1745-2_5.

[11] P. Montmitonnet, « Hot and cold strip rolling processes », *Computer Methods in Applied Mechanics and Engineering*, vol. 195, nº 48-49, p. 6604-6625, oct. 2006, doi: 10.1016/j.cma.2005.10.014.

[12] C. Counhaye, « Modélisation et contrôle industriel de la géométrie des aciers laminés à froid », Ph.D, Université de Liège, 2000.

[13] G.-J. Li et S. Kobayashi, « Rigid-Plastic Finite-Element Analysis of Plane Strain Rolling », *Journal of Engineering for Industry*, vol. 104, nº 1, p. 55-63, févr. 1982, doi: 10.1115/1.3185797.

[14] K. Mori et K. Osakada, « Simulation of three-dimensional deformation in rolling by the finite-element method », *International Journal of Mechanical Sciences*, vol. 26, nº 9-10, p. 515-525, janv. 1984, doi: 10.1016/0020-7403(84)90005-5.

[15] H. J. Huisman et J. Huetink, « A combined eulerian-lagrangian three-dimensional finite-element analysis of edge-rolling », *Journal of Mechanical Working Technology*, vol. 11, nº 3, p. 333-353, juill. 1985, doi: 10.1016/0378-3804(85)90005-1.

[16] J. L. Chenot, P. Montmitonnet, A. Bern, et C. Bertrand-Corsini, « A method for determining free surfaces in steady state finite element computations », *Computer Methods in Applied Mechanics and Engineering*, vol. 92, nº 2, p. 245-260, nov. 1991, doi: 10.1016/0045-7825(91)90242-X.

[17] T. Coupez, « Parallel adaptive remeshing in 3D moving mesh finite element », in *Report Number: CONF-960489--*, United States: Mississippi State Univ., Mississippi State, MS (United States), 30apr. J.-C. [En ligne]. Disponible sur: https://www.osti.gov/biblio/416600

[18] Y. Zhang, C.-A. Gandin, et M. Bellet, « Finite Element Modeling of Powder Bed Fusion at Part Scale by a Super-Layer Deposition Method Based on Level Set and Mesh Adaptation », *Journal of Manufacturing Science and Engineering*, vol. 144, nº 5, p. 051001, mai 2022, doi: 10.1115/1.4052386.

[19] R. Ducloux, L. Fourment, S. Marie, et D. Monnereau, « Automatic Optimization Techniques Applied to a Large Range of Industrial Test Cases », *Int J Mater Form*, vol. 3, nº S1, p. 53-56, avr. 2010, doi: 10.1007/s12289-010-0705-4.

[20] W. T. Lankford, S. C. Snyder, et J. A. Bausher, « New criteria for predicting the performance of deep drawing sheets », *Transactions of the American Society for Metals*, vol. 42, p. 1197, 1950.

[21] L. Maire, J. Fausty, M. Bernacki, N. Bozzolo, P. De Micheli, et C. Moussa, « A new topological approach for the mean field modeling of dynamic recrystallization », *Materials & Design*, vol. 146, p. 194-207, mai 2018, doi: 10.1016/j.matdes.2018.03.011.

[22] L. Maire *et al.*, « 3D Full field modelling of recrystallization in a finite element framework – application to 304L », mai 2017. Consulté le: 4 septembre 2018. [En ligne]. Disponible sur: https://hal-mines-paristech.archives-ouvertes.fr/hal-01521699/document

[23] T. C. Cao, « Modeling ductile damage for complex loading paths. », Ph.D, Mines Paris PSL, 2013.

[24] Y. Li, M. Luo, J. Gerlach, et T. Wierzbicki, « Prediction of shear-induced fracture in sheet metal forming », *Journal of Materials Processing Technology*, vol. 210, nº 14, p. 1858-1869, nov. 2010, doi: 10.1016/j.jmatprotec.2010.06.021.

[25] J. Cao, M. Bambach, M. Merklein, M. Mozaffar, et T. Xue, « Artificial intelligence in metal forming », *CIRP Annals*, vol. 73, nº 2, p. 561-587, 2024, doi: 10.1016/j.cirp.2024.04.102.

[26] J. F. Sacadura, *Transferts thermiques - Initiation et approfondissement*, Tec&Doc, Lavoisier. Paris, 2015.

[27] T. Wanheim, N. Bay, et A. S. Petersen, « A theoretically determined model for friction in metal working processes », *Wear*, vol. 28, p. 251-258, 1974, doi: 10.1016/0043-1648(74)90165-3.

[28] N. Bay, « Friction stress and normal stress in bulk metal-forming processes », *Journal of Mechanical Working Technology*, vol. 14, p. 203-223, 1987, doi: 10.1016/0378-3804(87)90061-1.

[29] P. Montmitonnet, « Lois de frottement », *Matériaux & Techniques*, vol. 81, nº 1-2-3, p. 8-21, 1993, doi: 10.1051/mattech/199381010008.

[30] J. F. Archard et W. Hirst, « The wear of metals under unlubricated conditions », *Proceedings of the Royal Society of London. Series A. Mathematical and Physical Sciences*, vol. 236, n° 1206, p. 397-410, janv. 1997, doi: 10.1098/rspa.1956.0144.

[31] M. Laugier, M. Tornicelli, C. S. Leligois, D. Bouquegneau, D. Launet, et J. A. Alvarez, « Flexible lubrication concept. The future of cold rolling lubrication », *Proceedings of the Institution of Mechanical Engineers, Part J: Journal of Engineering Tribology*, vol. 225, n° 9, p. 949-958, sept. 2011, doi: 10.1177/1350650111414514.

[32] M. Massin, « La lubrification des micro-mécanismes par les polymères silicones », *Polymères et lubrification*, vol. 233, p. 95, 1975.

[33] W. R. D. Wilson et J. A. Walowit, « An isothermal lubrication theory for strip rolling with front and back tension. Proc. Of Trib. Conv. 1971, Douglas, Isle of Man, UK, 12th-15th May 1971. Proc IMechE, 1972, pp. 164-172 », *Proceedings of Tribology Convention 1971*, p. 164-172, 1972.

[34] W. R. D. Wilson, « The temporary breakdown of hydrodynamic lubrication during the initiation of extrusion », *International Journal of Mechanical Sciences*, vol. 13, n° 1, p. 17-28, janv. 1971, doi: 10.1016/0020-7403(71)90099-3.

[35] W. R. D. Wilson, « An Isoviscous Model for the Hydrodynamic Lubrication of Plane Strain Forging Processes With Flat Dies », *Journal of Lubrication Technology*, vol. 96, n° 4, p. 539-546, oct. 1974, doi: 10.1115/1.3452478.

[36] W. R. D. Wilson et J. J. Wang, « Hydrodynamic Lubrication in Simple Stretch Forming Processes », *Journal of Tribology*, vol. 106, n° 1, p. 70-77, janv. 1984, doi: 10.1115/1.3260870.

[37] W. R. D. Wilson, Y. Sakaguchi, et S. R. Schmid, « A dynamic concentration model for lubrication with oil-in-water emulsions », *Wear*, vol. 161, n° 1-2, p. 207-212, avr. 1993, doi: 10.1016/0043-1648(93)90471-W.

[38] S. Cassarini, M. Laugier, et P. Montmitonnet, « Modelling of metal forming lubrication by o/w emulsions. Proc. 2nd Int. Conf. Trib. Manuf. Proc., Nyborg (Denmark), 2004, pp. 713-720 »,

[39] D. Tabor, « Mohs's Hardness Scale - A Physical Interpretation », *Proc. Phys. Soc. B*, vol. 67, n° 3, p. 249-257, mars 1954, doi: 10.1088/0370-1301/67/3/310.

[40] P. G. Stevens, K. P. Ivens, et P. Harper, « Increasing work-roll life by improved roll-cooling practice », *J Iron Steel Inst*, vol. 209, n° 1, p. 1-11, 1971.

[41] P. Montmitonnet, F. Delamare, et B. Rizoulières, « Couches de Transfert et Tribologie du Laminage », in *Surfaces, tribologie et formage des matériaux - Eric Felder - Librairie Eyrolles*, Les presse de l'école des Mines de Paris., 2001.

[42] P. Montmitonnet, F. Delamare, et B. Rizoulieres, « Transfer layer and friction in cold metal strip rolling processes », *Wear*, vol. 245, n° 1-2, p. 125-135, oct. 2000, doi: 10.1016/S0043-1648(00)00473-7.

[43] D. Hull et D. Bacon, *Introduction to dislocations*, Fourth edition. Butterworth-Heinemann, 2001.

[44] D. Irmer, C. Moussa, L. T. Belkacemi, M. Sennour, A. Vaissière, et V. A. Esin, « Effect of cold rolling on nucleation, growth and coarsening of S-phase precipitates in Al- Cu- Mg alloy (AA2024): From heterogeneous nucleation to homogeneous spatial distribution », *Journal of Alloys and Compounds*, vol. 963, p. 171162, nov. 2023, doi: 10.1016/j.jallcom.2023.171162.

[45] M. Saby, P.-O. Bouchard, et M. Bernacki, « Void closure criteria for hot metal forming: A review », *Journal of Manufacturing Processes*, vol. 19, p. 239-250, août 2015, doi: 10.1016/j.jmapro.2014.05.006.

[46] J. Humphreys, G. S. Rohrer, et A. Rollett, *Recrystallization and Related Annealing Phenomena (Third Edition)*. Oxford: Elsevier, 2017. doi: 10.1016/B978-0-08-098235-9.00001-X.

[47] U. F. Kocks et H. Mecking, « Physics and phenomenology of strain hardening: the FCC case », *Progress in Materials Science*, vol. 48, n° 3, p. 171-273, janv. 2003, doi: 10.1016/S0079-6425(02)00003-8.

[48] O. Engler et V. Randle, *Introduction to Texture Analysis: Macrotexture, Microtexture, and Orientation Mapping, Second Edition*, 0 éd. CRC Press, 2009. doi: 10.1201/9781420063660.

[49] U. F. Kocks, C. N. Tomé, et H. R. Wenk, *Texture and Anisotropy*, Cambridge University Press. 1998.

[50] R. Pokharel *et al.*, « Polycrystal Plasticity: Comparison Between Grain - Scale Observations of Deformation and Simulations », *Annual Review of Condensed Matter Physics*, vol. 5, nº 1, p. 317-346, 2014, doi: 10.1146/annurev-conmatphys-031113-133846.

[51] E. Nes, « Modelling of work hardening and stress saturation in FCC metals », *Progress in Materials Science*, vol. 41, nº 3, p. 129-193, juill. 1997, doi: 10.1016/S0079-6425(97)00032-7.

[52] D. Irmer, C. Yildirim, M. Sennour, V. A. Esin, et C. Moussa, « Effect of second-phase precipitates on deformation microstructure in AA2024 (Al–Cu–Mg): dislocation substructures and stored energy », *J Mater Sci*, vol. 59, nº 40, p. 18978-19002, oct. 2024, doi: 10.1007/s10853-024-10205-6.

[53] J. Baton, W. Geslin, et C. Moussa, « Orientation and deformation conditions dependence of dislocation substructures in cold deformed pure tantalum », *Materials Characterization*, vol. 171, p. 110789, janv. 2021, doi: 10.1016/j.matchar.2020.110789.

[54] N. Hansen, « New discoveries in deformed metals », *Metallurgical and Materials Transactions A*, vol. 32, nº 12, p. 2917-2935, déc. 2001, doi: 10.1007/s11661-001-0167-x.

[55] J. Baton, W. Geslin, et C. Moussa, « Influence of pre-recovery on the recrystallization of pure tantalum », *J Mater Sci*, vol. 56, nº 27, p. 15354-15378, sept. 2021, doi: 10.1007/s10853-021-06218-0.

[56] H. Fan, S. Liu, C. Deng, X. Wu, L. Cao, et Q. Liu, « Quantitative analysis: How annealing temperature influences recrystallization texture and grain shape in tantalum », *International Journal of Refractory Metals and Hard Materials*, vol. 72, p. 244-252, avr. 2018, doi: 10.1016/j.ijrmhm.2017.12.003.

[57] W. T. Read et W. Shockley, « Dislocation models of crystal grain boundaries », *Physical review*, vol. 78, p. 275, 1950.

[58] A. Godfrey, W. Q. Cao, Q. Liu, et N. Hansen, « Stored energy, microstructure, and flow stress of deformed metals », *Metall Mater Trans A*, vol. 36, nº 9, p. 2371-2378, sept. 2005, doi: 10.1007/s11661-005-0109-0.

[59] B. Derby, « The dependence of grain size on stress during dynamic recrystallisation », *Acta Metallurgica et Materialia*, vol. 39, nº 5, p. 955-962, mai 1991, doi: 10.1016/0956-7151(91)90295-C.

[60] K. Huang et R. E. Logé, « A review of dynamic recrystallization phenomena in metallic materials », *Materials & Design*, vol. 111, p. 548-574, déc. 2016, doi: 10.1016/j.matdes.2016.09.012.

[61] F. Grégori, K. Murakami, et B. Bacroix, « The influence of microstructural features of individual grains on texture formation by strain-induced boundary migration in non-oriented electrical steels », *J Mater Sci*, vol. 49, nº 4, p. 1764-1775, févr. 2014, doi: 10.1007/s10853-013-7863-y.

[62] C. Xu *et al.*, « Direct observation of nucleation in the bulk of an opaque sample », *Scientific Reports*, vol. 7, p. 42508, févr. 2017.

[63] J. Zhang *et al.*, « Grain boundary mobilities in polycrystals », *Acta Materialia*, vol. 191, p. 211-220, juin 2020, doi: 10.1016/j.actamat.2020.03.044.

[64] M. Bernacki, « Kinetic equations and level-set approach for simulating solid-state microstructure evolutions at the mesoscopic scale: State of the art, limitations, and prospects », *Progress in Materials Science*, vol. 142, p. 101224, avr. 2024, doi: 10.1016/j.pmatsci.2023.101224.

[65] J. W. Christian, *The Theory of Transformations in Metals and Alloys*. Elsevier, 2002. doi: 10.1016/B978-0-08-044019-4.X5000-4.

[66] D. A. Porter, K. E. Easterling, et M. Y. Sherif, *Phase Transformations in Metals and Alloys*, 4ᵉ éd. Boca Raton: CRC Press, 2021. doi: 10.1201/9781003011804.

[67] C. Moussa *et al.*, « Microstructural Evolution in Inconel 718 During Room-Temperature Tube Flow Forming: Plastic Flow Heterogeneities and δ Phase Evolution », *Metall Mater Trans A*, vol. 55, nº 5, p. 1311-1318, mai 2024, doi: 10.1007/s11661-024-07361-4.

[68] K. Zheng *et al.*, « The effect of hot form quench (HFQ®) conditions on precipitation and mechanical properties of aluminium alloys », *Materials Science and Engineering: A*, vol. 761, p. 138017, juill. 2019, doi: 10.1016/j.msea.2019.06.027.